\documentclass{article}[10pt]
\usepackage{algorithmicx}
\usepackage{algorithm}
\usepackage{amsmath,amssymb}

\usepackage{algpseudocode}
\usepackage{multirow}
\usepackage{epsfig}
\usepackage{multicol,lipsum,xparse}
\usepackage{subcaption}
\usepackage{amsmath}
\usepackage{stmaryrd}
\usepackage{color}
\usepackage{todonotes}

\begin{document}

\title{Algorithmic Obfuscation over GF({\LARGE $2^m$})}

\author{Cunxi Yu$^1$, Daniel Holcomb$^2$\\
Cornell University $^1$\\
University of Massachusetts Amherst$^2$ \\
cunxi.yu@cornell.edu}


%
\maketitle

\vspace{-5mm}

{\it\bf Abstract -} 


Galois Field arithmetic blocks are the key components in many security applications, such as \textit{Elliptic Curve Cryptography} (ECC) and the S-Boxes of the \textit{Advanced Encryption Standard} (AES) cipher. This paper introduces a novel hardware intellectual property (IP) protection technique by obfuscating arithmetic functions over Galois Field (GF), specifically, focusing on obfuscation of GF multiplication that underpins complex GF arithmetic and \textit{elliptic curve point} arithmetic functions. Obfuscating GF multiplication circuits is important because the choice of irreducible polynomials in GF multiplication has the great impact on the performance of the hardware designs, and because the significant effort is spent on finding an optimum irreducible polynomial for a given field, which can provide one company a competitive advantage over another. 


\section{Introduction}

Due to the increasing cost of integrated circuit (IC) design and manufacturing, it becomes more important to protect the intellectual property (IP) of an IC against reverse engineering (RE). Despite the complexity of modern ICs, the implementation details can be extracted by RE techniques once a circuit is fabricated and released to market~\cite{TorranceJ11}. This has given rise to a number of academic works and commercial products focused on using obfuscation to make designs more difficult to reverse engineer. In this work, we focus on protecting the IP of Galois Field arithmetic circuits that are commonly used in cryptography.

Galois field (GF) is a number system with a finite number of elements. Galois Field arithmetic is used extensively in many security applications such as \textit{Elliptic Curve Cryptography} (ECC) and the \textit{Advanced Encryption Standard} (AES). 
The basic arithmetic functions include GF addition and multiplication, and more advanced GF arithmetic functions are derived from those two \cite{paar2009understanding}, such as GF division, and \textit{elliptic curve} point addition and multiplication \cite{koblitz1987elliptic}. An irreducible polynomial $P(x)$ is required for constructing GF multiplication, and the choice of polynomial has a significant impact on the performance of the GF multiplier. 

Intel Research showed that GF($2^4$) (4-bit) GF multipliers implemented in the same 22nm CMOS technology with two different $P(x)$ differ by 40\% and 9\% in area and delay respectively, on their \textit{nanoAES} chip\cite{mathew2015340}. The costs vary with the choice of polynomial because different polynomials require different numbers of XOR operations in the critical path and the entire design; as the bit-width increases, the performance of the multipliers varies even more strongly across different polynomials \cite{YuDATE16ReverseEng}. Moreover,  recent work shows that it is possible to reverse engineer the irreducible polynomial of a post-synthesized GF($2^m$) multiplier up to 571-bit~\cite{YuDATE16ReverseEng} by analyzing the algebraic signatures extracted from the gate-level netlist. This shows that the choice of polynomial should not be considered secret by default unless steps are taken to obfuscate it. Due to the significant efforts spent on finding the optimum $P(x)$ in industrial designs~\cite{mathew2015340}, preventing the $P(x)$ from being reverse engineered becomes important\footnote{We specifically refer to Section IV in \cite{mathew2015340} and Section II-D in \cite{YuDATE16ReverseEng}.}.


This work introduces the first obfuscation methodology over Galois Field that obfuscates the GF multiplication with multiple irreducible polynomials, and furthermore shows that there is a small performance overhead per obfuscated function. The obfuscation approach is based on analyzing the mathematical properties of finite field arithmetic to identify the maximum common logic between multiplications of different $P(x)$. The resulting multiplier performs multiplication with a certain $P(x)$, denoted the \textit{true function}, which is known by the designers only. An attacker should be unable to use reverse engineering to distinguish the true function from the obfuscated functions which are all valid GF multiplications in the same finite field as the true function.


\section{Background} \label{sec:background}

\subsection{Galois Field Principle}\label{sec:background:gf}

Galois field (GF) is a number system with a finite number of elements and two main arithmetic operations, addition and multiplication; other operations such as division can be derived from those two \cite{paar2009understanding}. Galois field with $p$ elements is denoted as GF($p$). Prime field, denoted GF($p$), is a finite field consisting of a finite number of integers \{$1,2, ....,p-1$\}, where $p$ is a prime number, with additions and multiplication performed \textit{modulo p}. 
Binary extension field, denoted GF($2^m$) (or $\mathbb{F}_{2^m}$), is a finite field with $2^m$ elements. Unlike in prime fields, however, the operations in extension fields are not computed \textit{modulo $2^{m}$}. Instead, in one possible representation (called \textit{polynomial basis}), each element of GF($2^m$) is a {\it polynomial ring} with $m$ terms with coefficients in GF(2) and modulo an irreducible polynomial $P(x)$. The addition of finite field elements is the addition of polynomials, with coefficients computed in GF(2).
Multiplication of field elements is performed modulo {\it irreducible polynomial} $P(x)$ of degree $m$ and coefficients in GF(2). The irreducible polynomial $P(x)$ is analogous to the prime number $p$ in prime fields $GF(p)$. 
Extension fields are used in many cryptography applications, such as AES and ECC.

\begin{figure}[!htb]
\scriptsize
\centering
\begin{tabular}{lllllll}
             &              &              & $a_3$        & $a_2$        & $a_1$        & $a_0$        \\
             &              &              & $b_3$        & $b_2$        & $b_1$        & $b_0$        \\ \hline
             &              &              & $a_{3}b_{0}$ & $a_{2}b_{0}$ & $a_{1}b_{0}$ & $a_{0}b_{0}$ \\
             &              & $a_{3}b_{1}$ & $a_{2}b_{1}$ & $a_{1}b_{1}$ & $a_{0}b_{1}$ &              \\
             & $a_{3}b_{2}$ & $a_{2}b_{2}$ & $a_{1}b_{2}$ & $a_{0}b_{2}$ &              &              \\
$a_{3}b_{3}$ & $a_{2}b_{3}$ & $a_{1}b_{3}$ & $a_{0}b_{3}$ &              &              &              \\ \hline
$s_6$        & $s_5$        & $s_4$        & $s_3$        & $s_2$        & $s_1$        & $s_0$      \vspace{2mm} \\
\multicolumn{7}{c}{$s_q$ = $\bigoplus a_i b_j$, $\forall$ $i$+$j$=$q$,  $0 \leq q \leq 6$}
\vspace{2mm}
\end{tabular}
\label{my-label}
        \begin{minipage}{.45\linewidth}
      \centering
\begin{tabular}{cccc}
\multicolumn{1}{c|}{$s_3$} & \multicolumn{1}{c|}{$s_2$} & \multicolumn{1}{c|}{$s_1$} & $s_0$                     \\
\multicolumn{1}{c|}{$s_4$} & \multicolumn{1}{c|}{0}  & \multicolumn{1}{c|}{0} & $s_4$                     \\
\multicolumn{1}{c|}{$s_5$} & \multicolumn{1}{c|}{0}  & \multicolumn{1}{c|}{$s_5$}  & $s_5$                     \\
\multicolumn{1}{c|}{$s_6$} & \multicolumn{1}{c|}{$s_6$} & \multicolumn{1}{c|}{$s_6$}  & $s_6$                      \\ \hline
\multicolumn{1}{l|}{$z_3$} & \multicolumn{1}{l|}{$z_2$} & \multicolumn{1}{l|}{$z_1$} & \multicolumn{1}{l}{$z_0$}
\end{tabular}
    \end{minipage}%
    \begin{minipage}{.55\linewidth}
      \centering
\begin{tabular}{l}
\multicolumn{1}{l}{$z_0$ = $s_0$ $\oplus$ $s_4$ $\oplus$ $s_5$ $\oplus$ $s_6$}              \\       
\multicolumn{1}{l}{$z_1$ = $s_1$ $\oplus$ $s_5$ $\oplus$ $s_6$}                     \\
\multicolumn{1}{l}{$z_2$ = $s_2$ $\oplus$ $s_6$}                     \\
\multicolumn{1}{l}{$z_3$ = $s_3$ $\oplus$ $s_4$ $\oplus$ $s_5$ $\oplus$ $s_6$}                 
\end{tabular}    \end{minipage} 
\caption{GF($2^4$) multiplication $Z$ mod $P(x)$ = $A\cdot B$ mod $P(x)$, where $P(x)$=$x^{4}+x^{3}+1$. $A$=$a_3 x^3$+$a_2 x^2$+$a_1 x^1$+$a_0$, $B$=$b_3 x^3$+$b_2 x^2$+$b_1 x^1$+$b_0$, etc. $Z$ are the output of the multiplication.}
\vspace{-2mm}
\label{fig:4-bit-gf}
\end{figure}


In an extension field with $p=2$, a polynomial representation is alternatively represented by a unique binary expression. This binary expression is the coefficients of each term in the polynomial. For example, the binary expression of $x^3$+$x^2$+$x^1$ is [$1110$]. The linear arithmetic operations in the finite field, i.e., addition and subtraction over GF($2^m$), are performed by adding or subtracting two of those polynomials. The result is reduced by modulo $p$, where $p=2$ in GF($2^m$). {In such a finite field, addition, subtraction modulo 2, and bit-vector XOR perform the same function. Thus, GF addition and subtraction at Boolean level are implemented with bit-vector XOR operations.} 
For example, ($x^3$+$x^2$+$x$)+($x^3$+$x^2$+$1$) = $2x^3$+$2x^2$+$x$+$1$ $mod~2$ = $x$+$1$, which is performed by [$1110$] $\oplus$ [$1101$] = [$0011$], which represents the polynomial $x$+$1$. 

Galois Field multiplication is performed by multiplication modulo an irreducible polynomial that defines the finite field. An irreducible polynomial is a polynomial that cannot be factored into nontrivial polynomials over the same field \cite{nagell1951introduction}. For example, in GF(2), $x^2$+$x$+$1$ is an irreducible polynomial but $x^2$+1 is not, since $x^2$+1=($x$+$1$)($x$+$1$). An example of GF($2^4$) (4-bit) multiplication is shown in Figure \ref{fig:4-bit-gf}, with irreducible polynomial $P(x)$=$x^4$+$x^3$+$1$. Similar to addition and subtraction, the inputs and outputs in multiplication are binary expressions. For example, A = [$a_3$ $a_2$ $a_1$ $a_0$], where $a_0$ is the least significant bit and $a_3$ is the most significant bit. The multiplication is performed in two stages: 1) adding the partial products and 2) reducing over GF($2^4$) with $P(x)$. The partial products are generated similarly to the integer multiplication using AND operations. Since additions in the field are XOR operations, the sum of the partial products ($s_{q}$ in Figure \ref{fig:4-bit-gf}) is generated using a series of XORs. 

The sum of partial products will then be reduced modulo the irreducible polynomial. As mentioned previously, the binary expression corresponds to the coefficients of polynomial expression. Thus, $s_i$ is the coefficient of $x^i$ in its polynomial representation. According to the modulo addition rule, {$s_0$+$s_1$+...$s_6$ mod $P(x)$ = [($s_0$ mod $P(x)$)+($s_1$ mod $P(x)$)+...($s_6$ mod $P(x)$)] mod $P(x)$.} The GF multiplication can be constructed as follows:

\begin{itemize}
\item $s_{2}$$x^0$ ~mod $P(x)$ = $s_0$;  ~~~~~$s_{1}$$x$ mod $P(x)$ = $s_1$$x$;  \\$s_{2}$$x^2$ mod $P(x)$ = $s_1$$x^2$;  ~~~$s_{3}$$x^3$ mod $P(x)$ = $s_3$$x^3$.\\ Hence, $\sum_{i=0}^{3} s_i$ mod $P(x)$ = $s_0$+$s_{1}$$x$+$s_{2}$$x^2$+$s_{3}$$x^3$, denoted as $P_{0}$.

\item $P_{1}$ = $s_{4}$$x^4$ mod $P(x)$ = $s_{4}$+$s_4$$x^3$ mod P(x).

\item $P_{2}$ = $s_{5}$$x^5$ mod $P(x)$ = $s_{5}$+$s_5$$x$+$s_4$$x^3$ mod P(x).

\item $P_{3}$ = $s_{6}$$x^6$ mod $P(x)$ = $s_6$+$s_{6}$$x$+$s_{6}$$x^2$+$s_{6}$$x^3$ mod P(x).

\item Hence, $\sum_{i=0}^{6} s_i$ mod $P(x)$ = $P_{0}$+$P_{1}$+$P_{2}$+$P_{3}$ mod $P(x)$, which is performed by GF additions.

\end{itemize}

Since GF additions are the same as XOR in the binary expressions of the polynomials, the final results [$z_3$ $z_2$ $z_1$ $z_0$] are produced by:
[$s_3$ $s_2$ $s_1$ $s_0$]$_{P_{0}}$ $\oplus$ [$s_4$ $0$ $0$ $s_0$]$_{P_{1}}$ $\oplus$ [$s_5$ $0$ $s_5$ $s_5$]$_{P_{2}}$ $\oplus$ [$s_6$ $s_6$ $s_6$ $s_6$]$_{P_{3}}$, as shown in Figure \ref{fig:4-bit-gf}. Galois Field addition and multiplication are the basic GF arithmetic operations that are used to implement the advanced arithmetic functions such as division, and \textit{elliptic-curve} point addition and multiplication \cite{koblitz1987elliptic} for cryptography applications. {However, GF addition is performed with one bit-vector XOR regardless of the irreducible polynomial, which means that obfuscating multiple irreducible polynomials cannot be applied to a stand-alone GF adder.} Thus, this work focuses on multiplication obfuscation over GF($2^m$).

\vspace{-2mm}
\subsection{Irreducible Polynomials}\label{sec:irreducible_poly}

\begin{table}[!htb]
\centering
\caption{Irreducible polynomials with degree $m$.}
\label{tbl:2_to_4_px}
\begin{tabular}{|l|l|}
\hline
\textit{m} & Irreducible polynomial(s) \\ \hline
2 & $x^2$+$x$+1 \\ \hline
3 & $x^3$+$x$+1; $x^3$+$x^2$+1 \\ \hline
4 & $x^4$+$x$+1; $x^4$+$x^3$+1; $x^4$+$x^3$+$x^2$+$x^1$+1 \\ \hline
5 & \begin{tabular}[c]{@{}l@{}}$x^5$+$x^2$+1; $x^5$+$x^3$+$x^2$+1; $x^5$+$x^3$+1; \\ $x^5$+$x^4$+$x^3$+1; $x^5$+$x^4$+$x^3$+$x^2$+1; $x^5$+$x^4$+$x^2$+$x$+1 \end{tabular} \\ \hline
\end{tabular}
\end{table}

In general, there are various irreducible polynomials that can be used for a given field size, each resulting in a different multiplication result. {The number of irreducible polynomials increases as $m$ increasing.} The list of irreducible polynomials that exist for degrees $m$=\{2,3,4,5\} are shown in Table \ref{tbl:2_to_4_px}. For constructing efficient arithmetic functions over GF($2^m$), the irreducible polynomial is typically chosen to be a trinomial, $x^m$+$x^a$+1, or a pentanomial $x^m$+$x^a$+$x^b$+$x^c$+1 \cite{nist-recommend}. It is furthermore required that coefficients $m, ~a$ be chosen such that $m$ - $a$ $\geq$ $m/2$ \cite{scott2007optimal}. 

\vspace{-2mm}
\begin{figure}[!htb]
\centering
\includegraphics[width=0.6\textwidth]{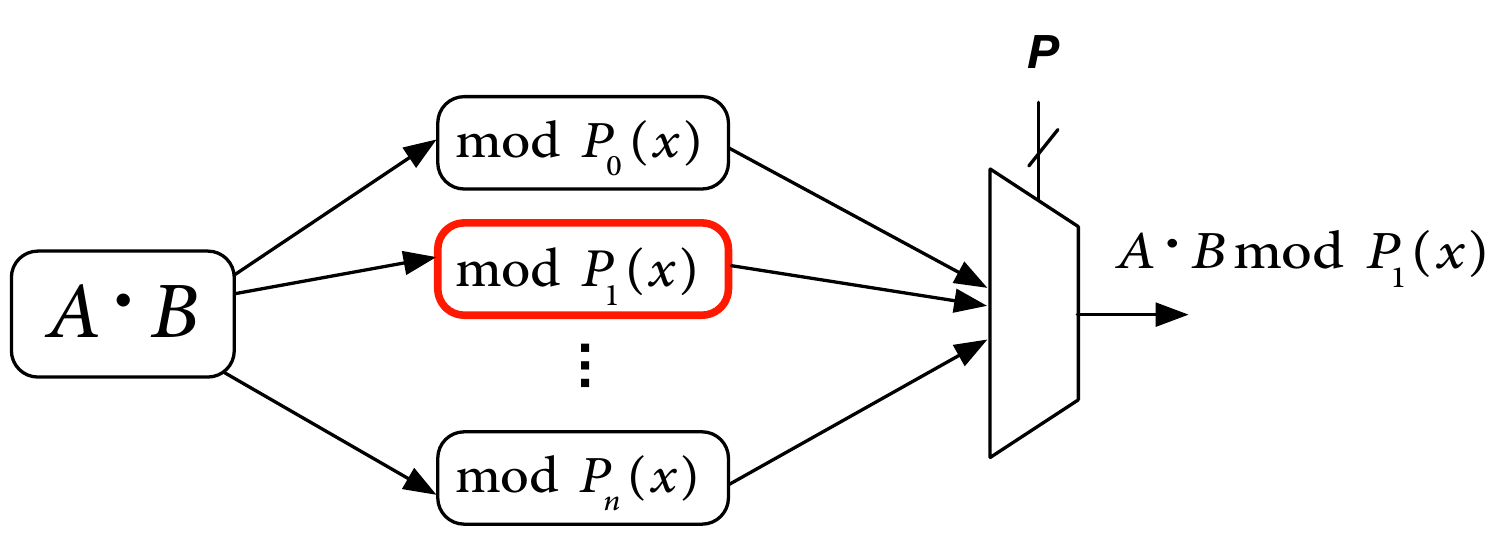}
\caption{Illustrative example of implementing GF multiplication with irreducible polynomial obfuscated.}
\vspace{-2mm}
\label{fig:motivation}
\end{figure}


Given degree $m$, the multiplications constructed by different irreducible polynomials are {functionally} different but are in the same field. For example, given degree $m$=4, in contrast to using $P_{0}(x)$=$x^4$+$x^3$+1, using $P_{1}(x)$=$x^4$+$x$+1 produces a different GF multiplication function in GF($2^4$). The difference appears in the process of reducing the sum of the partial products modulo the irreducible polynomial. For example, when using polynomial $P_{0}$, $s^6$ is required for all the output bits since $x^6$ mod $P_{0}(x)$ is equal to $x^3$+$x^2$+$x^1$+$1$, which needs \underline{four} XOR operations. However, when using polynomial $P_{1}(x)$, $s^6$ is only required for $z_2$ and $z_3$ since $x^6$ mod $P_{1}(x)$ is equal to $x^3$+$x^2$, which needs only \underline{two} XOR operations. This explains why the choice of irreducible polynomials effects the delay and area of GF multipliers.


The goal of our approach is implementing a multiplier with multiple functions obfuscated, such that only designers and authorized users know the true function (Figure \ref{fig:motivation}). $P$ are the switches that will be physically implemented as constant inputs using \textit{camouflaged standard cells}. One important observation is that most logic of the GF multipliers using different irreducible polynomials remain the same. The main differences are the logic of reducing the sum of the partial products. This means that the overhead of implementing such an obfuscated GF($2^m$) multiplier is much smaller than synthesizing the model in Figure \ref{fig:motivation}, which offers the main motivation for this work. 

\vspace{-2mm}
\subsection{Camouflaged Standard Cell}\label{sec:background_cam_cell}

Gate-level camouflaging techniques rely on using camouflaged standard cells in the fabricated integrated circuits. The camouflaged standard cells are designed as independent standard cells in the technology library. Mostly, the camouflaged cells are used to introduce dummy functionalities. During technology mapping in the design flow, the original functionality of the circuit is camouflaged by partially mapping the circuit with the camouflaged standard cells. These camouflaged cells are designed by changing the layout of the cell with dummy contacts \cite{RajendranSSK13-ccs}, or by changing doping of the transistors \cite{DBLP:conf/ches/BeckerRPB13}. With such camouflaged cells, designing circuits with constant inputs camouflaged becomes possible. For example, a 2-NAND is proposed to implement camouflaged constant one/zero by modifying the doping of different transistors \cite{DBLP:conf/date/KeshavarzPH17}. {A variant of dopant-programmable cells is to build components in a dual-Vt process technology such that inferring the correct component functions would require identification of which devices use high and low thresholds~\cite{collantes-16}. To minimize cost, it is often desirable to protect a circuit by camouflaging only a small subset of the gates~\cite{rajendran-13}.} However, in emerging technologies, it can be more difficult to infer function from structure~\cite{bi2016emerging}, and a reverse engineer may thus need to consider all gates as camouflaged. An overview of physical mechanisms for obfuscation is given by Vijayakumar et al. \cite{vijayakumar-2017-physical}. In this work, the switches $P$ shown in Figure \ref{fig:motivation} are mapped with such camouflaged standard cells.

\subsection{Attacker Model}\label{sec:attack_back}

The attacher model in this work is similar to the attacker model for reverse engineering circuits with camouflaged gates, which is firstly given by Rajendran \cite{rajendran-12}. The logic function implemented by a camouflaged circuit should remain hard to discover when the attacker has knowledge of all non-camouflaged gates and can apply inputs to the circuit and observe outputs. Techniques from oracle-guided synthesis~\cite{jha2010oracle} have recently been used in SAT-based attacks to reverse engineer gate camouflaging~\cite{elmassad-15} and logic encryption~\cite{subramanyan-15}, and improved with incremental SAT solving \cite{LiuYZH16} and approximate deobfuscation by relaxing the conjunctions of SAT formulas \cite{shamsi2017appsat}. With the knowledge of capabilities and limitations of oracle-guided SAT-based attacks, there are several countermeasure techniques developed, such as introducing AND-tree \cite{DBLP:conf/iccad/LiSMZYJP16}, protecting the minterms of the specification \cite{DBLP:conf/iccad/YasinMSR16}, introducing dummy combinational loops \cite{shamsi2017cyclic}, etc. Moreover, formal verification technique based on computer algebraic methods \cite{ciesielski2015verification,yu-aspdac-17} is shown to be able to reverse engineer the irreducible polynomials while the GF circuit is considered as a black box and the encodings of primary inputs and outputs are unknown \cite{yu:date2018-reverse}. 

We assume that the attackers know all the irreducible polynomials of a given field GF($2^m$). The attackers also have access to the physical implementation. We assume that they can obtain the gate-level netlist and identify the camouflaged standard cells that implement the constant inputs using reverse engineering techniques. The attackers aim to find the true irreducible polynomial used in the multiplier blocks in a large design. Using this knowledge to reverse engineer the true irreducible polynomial, the attackers have to do the following: 1) the attackers have to guess the value of the camouflaged signals implemented with the camouflaged standard cells; 2) prove the equivalence between the implementation and the reverse engineered version. The results of evaluating the obfuscation strength are shown in Section \ref{sec:attack},  with SAT-based attack techniques, and the recently proposed Binary Decision Diagrams (BDDs) based approach \cite{xu2017novel}. 


\begin{figure}[!htb]
\centering
\includegraphics[width=0.6\textwidth]{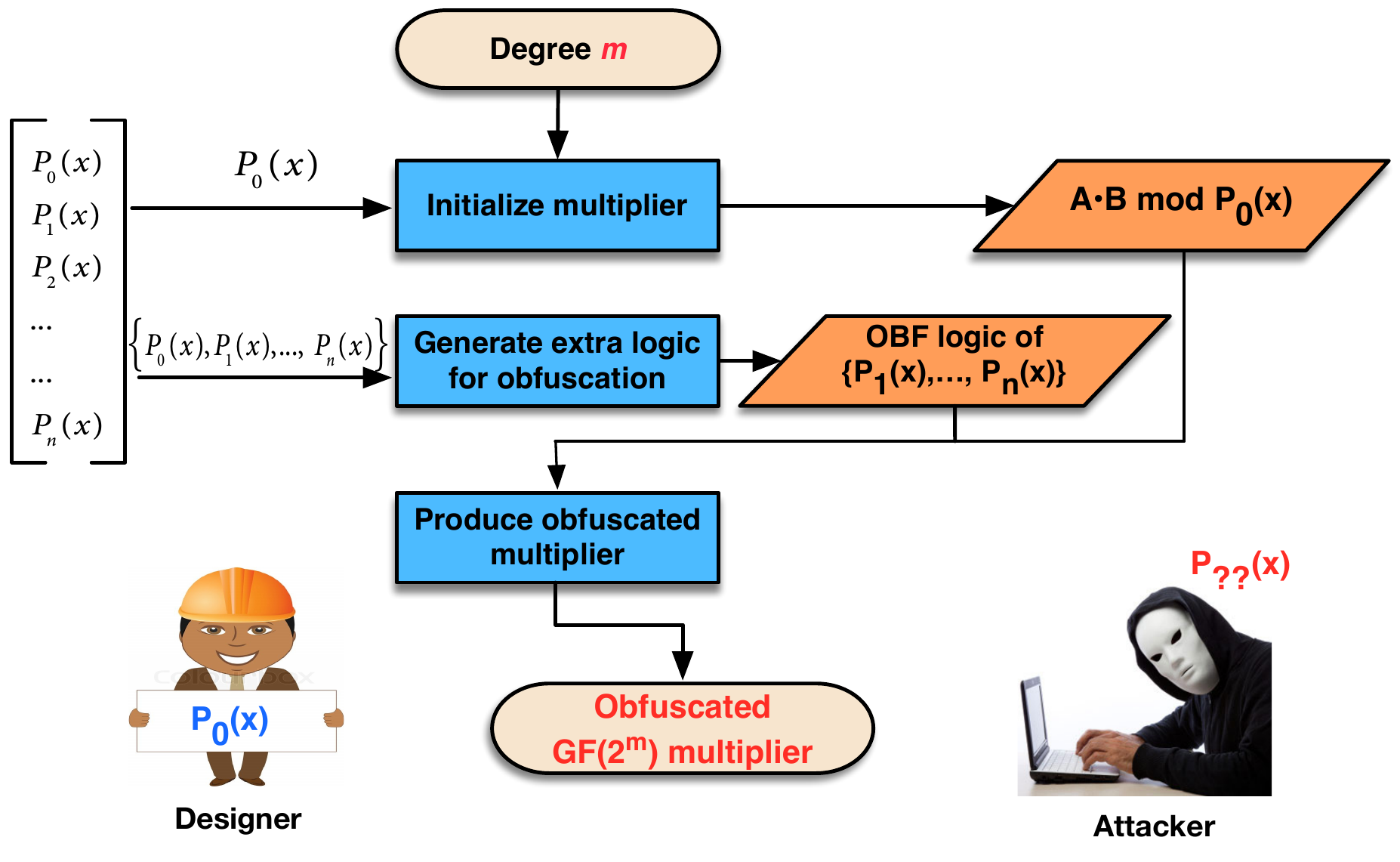}
\caption{Designing an obfuscated GF($2^m$) multiplier with $n+1$ functions.}
\vspace{-3mm}
\label{fig:overview}
\end{figure}

\section{Approach}

The flow of the proposed obfuscation approach is shown in Figure \ref{fig:overview}. The inputs include 1) the degree (bit-width) $m$ of the GF multiplier, 2) a set of irreducible polynomials in GF($2^m$), 3) the indication of the true function, and 4) the irreducible polynomials will be obfuscated in resulting multiplier. In Figure \ref{fig:overview}, the true function will be multiplication modulo $P_{0}(x)$, and multiplications modulo \{$P_{1}(x)$, $P_{2}(x)$, ..., $P_{n}(x)$\} are the obfuscated functions. The designer and authorized users know which irreducible polynomial is used for constructing the field. This approach is processed in three steps:

\begin{itemize}

\item \textbf{Initialize a GF($2^m$) multiplier with the irreducible polynomial that the designer wants to implement in the design.} This requires a function that generates the multiplication structure (e.g., Figure \ref{fig:4-bit-gf}) with any irreducible polynomial. Since the partial products and the sum of partial products are identical for the irreducible polynomials with the same degree, this function is reduced to produce the structure of reducing the addition of the partial products. 

\item \textbf{Generate and minimize the extra logic for adding obfuscated functions, i.e., multiplications modulo \{$P_{1}(x)$, $P_{2}(x)$, ..., $P_{n}(x)$\}.} Based on our observation, the only changes needed to add obfuscated functions are in the logic that reduces the sum of partial products modulo different irreducible polynomials. Thus, these logic can be produced by comparing the reduction structures. The output is an updated reduction structure. This function is applied iteratively for generating the obfuscation logic for $n$ polynomials.

\item \textbf{Produce the obfuscated multiplier}. This process generates the obfuscated multiplier by combining the partial product generator, the addition of partial products, and the reduction structure created by the previous step. The output of this process will be the input of the design flow, which produces the gate-level netlist and layouts. 

\end{itemize}

\begin{figure*}[t]
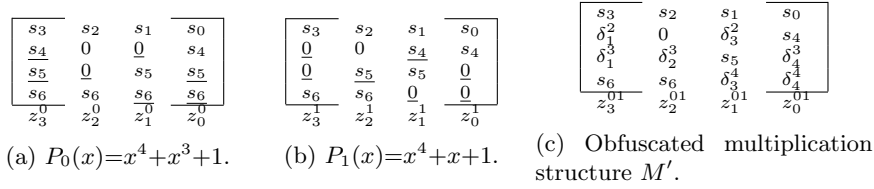

\scriptsize
\centering
        \begin{minipage}{.3\linewidth}
      \centering
\begin{tabular}{llll}
\cline{1-1} \cline{4-4}
\multicolumn{1}{|l}{$s_3$} & $s_2$ & $s_1$ & \multicolumn{1}{l|}{$s_0$} \\
\multicolumn{1}{|l}{\underline{$s_4$}}  & {0}  & \underline{0}  & \multicolumn{1}{l|}{$s_4$}  \\
\multicolumn{1}{|l}{\underline{$s_5$}} & \underline{0}  & $s_5$  & \multicolumn{1}{l|}{\underline{$s_5$}}  \\
\multicolumn{1}{|l}{$s_6$} & $s_6$  & \underline{$s_6$}  & \multicolumn{1}{l|}{\underline{$s_6$}} \\ \cline{1-1} \cline{4-4} 
$z^{0}_3$                      & $z^{0}_2$ & $z^{0}_1$ & $z^{0}_0$ 
\end{tabular}
\subcaption{$P_{0}(x)$=$x^4$+$x^3$+1.}
    \end{minipage}%
    \begin{minipage}{.3\linewidth}
      \centering
\begin{tabular}{llll}
\cline{1-1} \cline{4-4}
\multicolumn{1}{|l}{$s_3$} & $s_2$ & $s_1$ & \multicolumn{1}{l|}{$s_0$} \\
\multicolumn{1}{|l}{\underline{$0$}}  & 0  & \underline{$s_4$}  & \multicolumn{1}{l|}{$s_4$}  \\
\multicolumn{1}{|l}{\underline{$0$}} & \underline{$s_5$}  & $s_5$  &  \multicolumn{1}{l|}{\underline{0}} \\
\multicolumn{1}{|l}{$s_6$} & $s_6$  & \underline{0}  & \multicolumn{1}{l|}{\underline{0}}  \\ \cline{1-1} \cline{4-4} 
$z^{1}_3$                      & $z^{1}_2$ & $z^{1}_1$ & $z^{1}_0$ 
\end{tabular}
\subcaption{$P_{1}(x)$=$x^4$+$x$+1.}
    \end{minipage}
        \begin{minipage}{.37\linewidth}
      \centering
\begin{tabular}{cccc}
\begin{tabular}{llll}
\cline{1-1} \cline{4-4}
\multicolumn{1}{|l}{$s_3$} & $s_2$ & $s_1$ & \multicolumn{1}{l|}{$s_0$} \\
\multicolumn{1}{|l}{$\delta ^{2}_1$}  & {0}  & $\delta ^{2}_3$  & \multicolumn{1}{l|}{$s_4$}  \\
\multicolumn{1}{|l}{$\delta ^{3}_1$} & $\delta ^{3}_2$  & $s_5$  & \multicolumn{1}{l|}{$\delta ^{3}_4$}  \\
\multicolumn{1}{|l}{$s_6$}& $s_6$  & $\delta ^{4}_3$  & \multicolumn{1}{l|}{$\delta ^{4}_4$}  \\ \cline{1-1} \cline{4-4} 
$z^{01}_3$                      & $z^{01}_2$ & $z^{01}_1$ & $z^{01}_0$ 
\end{tabular}
\end{tabular}
\subcaption{Obfuscated multiplication structure $M'$.}
    \end{minipage}
\caption{\textbf{a, b)} GF($2^4$) multiplication structures $M_0$,$M_1$, with irreducible polynomials $P_{0}$ and $P_{1}$, generated by \textit{structureGen}; \textbf{c)} obfuscated multiplication structure, in which the true function is implemented with $P_{0}$.}
\label{fig:obf_4_example}
\end{figure*}


\begin{algorithm}[!htb]
\scriptsize
\caption{{\it structureGen(P(x))}: Generate Mult Structure}\label{alg:structureGen}
\textbf{Input: Irreducible polynomial $P(x)$ with degree $m$}\\ 
\textbf{Output: Galois Field Multiplication Structure}
\begin{algorithmic}[1]
\For{all $r$ and $c$, 0 $\leq$ $r,c$ $\leq$ m-1}
\State{$pp_{r,c}$ = $a_{r} \cdot b_{c}$}
\EndFor 
%
\For{$i$=0;$i$ $\leq$ 2m-2; $i$++}
\For{each $pp_{r,c}$, 0 $\leq$ $r,c$ $\leq$ m-1}
\If{$i$ == $r$+$c$}
\State $s_{i}$ = $s_{i}$ $\oplus$ $pp_{r,c}$
\EndIf
\EndFor
\EndFor 
\State Initialize an $m$-by-$m$ matrix $M$
\For{$i_1$=0; $i_1 \leq m-1$, $i_1$++}
\State $M(1,i_1)$ = $s_{(m-1) - i_{1}}$
\EndFor
\For{$i_2$=m; $i_2 \leq 2m-2$, $i_2$++}
\State {$\mathbb{R}$ mod $P(x)$ = $x^{i_2}$ mod $P(x)$}
\If{$x^q$ exists in $\mathbb{R}$, 0 $\leq$ q $\leq$m-1}
\State $M(i_{2}-m, q)$=$s_{i_2}$
\EndIf
\EndFor \\
\Return \{$pp_{r,c}$, 0 $\leq$ $r,c$ $\leq$ m-1 \}, \{$s_{i}$, 0 $\leq$ $i$ $\leq$ 2m-2 \} and $M$
\end{algorithmic}
\end{algorithm}

\subsection{Multiplication Structure Generation}

The algorithm of generating GF multiplication structure is shown in Algorithm \ref{alg:structureGen}, including \textit{partial products}, \textit{addition of partial products}, and \textit{reduction structure}. Algorithm \ref{alg:structureGen} is illustrated using another multiplication of GF($2^4$) using polynomial $x^4$+$x$+1. The first two functions are trivial and produce the same result as multiplication using $x^4$+$x^3$+1 (see Figure \ref{fig:4-bit-gf}).

The reduction structure is modeled as a matrix in this work. Thus, an $m$-by-$m$ matrix will initialized with all 0s (line 11). The first row of the matrix is filled with \{$s_{i}$, $0$ $\leq$ $i$ $\leq$ $m$-$1$\}. Note that the polynomial modulo function is not applied to those terms. This is because the result of $x^i$ modulo a polynomial with degree $m$ is always $x^i$ if $i$$<$$m$ is true (lines 2-4). To determine the rest of the structure, \{$x^i$, $m$ $\leq$ $i$ $\leq$ $2m$-$2$\} modulo $P(x)$ is then processed (line 16). The results are shown in Equations \ref{modulo_eqn1}-\ref{modulo_eqn3}. The $r^{th}$ row of the matrix (2 $\leq$ $r$ $\leq$m) is filled with $s_{m-2+r}$ by checking the existing terms in the remainder $\mathbb{R}$ of $x^i$ mod $x^4$+$x$+1 (lines 7-9). 

{\small
\vspace{-3mm}
\begin{equation} \label{modulo_eqn1}
\raggedleft
x^4~mod~x^4+x+1 \equiv x+1~mod~x^4+x+1
\end{equation}
\vspace{-6mm}
\begin{equation} \label{modulo_eqn2}
\raggedleft
x^5~mod~x^4+x+1 \equiv x^2+x~mod~x^4+x+1 
\end{equation}
\vspace{-6mm}
\begin{equation} \label{modulo_eqn3}
\raggedleft
x^6~mod~x^4+x+1 \equiv x^3+x^2~mod~x^4+x+1
\end{equation}
}
For example (Eq. \ref{modulo_eqn1}-\ref{modulo_eqn3}), according to lines 7-9 in Algorithm \ref{alg:structureGen}, $q^{th}$ position at $2^{nd}$ row is filled with $s_4$ if $x^q$ appears in the remainder. In Equation \ref{modulo_eqn1}, $x^1$ and $x^0$ appear in the remainder. Thus, $M(2,0)$ and $M(2,1)$ are filled with $s_4$. Similarly, $M(3,1)$ and $M(3,2)$ are filled with $s_5$, and $M(4,2)$ and $M(4,3)$ are filled with $s_6$. The rest of the elements in $M$ remain as 0. Applying Algorithm \ref{alg:structureGen} with two irreducible polynomials $P_0(x)$ and $P_1(x)$ over GF($2^4$) produces the results denoted as $M_{0}$ and $M_{1}$ in Figure \ref{fig:obf_4_example} (a) and (b)\footnote{The partial products and sum of the partial products structures are not included in this figure.}. 

\begin{table*}[]
\scriptsize
\centering
\caption{Evaluation of our obfuscation approach for GF($2^m$) multipliers where $m$=8, 64, 128, and 256, with 4, 8, and 16 functions obfuscated. *The area and delay results have been normalized.}
\label{tbl:all_results}
\begin{tabular}{|r|r|r|r|r|r|r|r|r|}
\hline
\multicolumn{1}{|c|}{\multirow{2}{*}{\begin{tabular}[c]{@{}c@{}}\# of\\ Functions\end{tabular}}} & \multicolumn{2}{c|}{\textit{GF($2^{8}$)}} & \multicolumn{2}{c|}{\textit{GF($2^{64}$)}} & \multicolumn{2}{c|}{\textit{GF($2^{128}$)}} & \multicolumn{2}{c|}{\textit{GF($2^{256}$)}} \\ \cline{2-9} 
\multicolumn{1}{|c|}{} & \multicolumn{1}{c|}{Area} & \multicolumn{1}{c|}{Delay} & \multicolumn{1}{c|}{Area} & \multicolumn{1}{c|}{Delay} & \multicolumn{1}{c|}{Area} & \multicolumn{1}{c|}{Delay} & \multicolumn{1}{c|}{Area} & \multicolumn{1}{c|}{Delay} \\ \hline
1 & 971 & 111.13 & 4.74e+04 & 175.01 & 1.89e+05 & 198.24 & 7.32e+05 & 224.14 \\ \hline
4 & 1735 & 139.07 & 5.46e+04 & 208.25 & 1.97e+05 & 246.65 & 7.57e+05 & 272.10 \\ \hline
8 & 2786 & 148.64 & 6.20e+04 & 225.33 & 2.13e+05 & 250.31 & 7.90e+05 & 274.92 \\ \hline
16 & \multicolumn{1}{c|}{-} & \multicolumn{1}{c|}{-} & 7.60e+04 & 240.09 & 2.48e+05 & 259.37 & 8.50e+05 & 281.87 \\ \hline
\end{tabular}
\vspace{-2mm}
\end{table*}

\vspace{-2mm}
\subsection{Obfuscation}\label{sec:obfuscation}
This section introduces our approach to generating the \textit{obfuscation logic} for producing the obfuscated multiplier. This procedure identifies the different logic between the original \underline{reduction structures}, and producing a new one with minimum extra logic introduced. The notations used in this section are as follows:

\begin{itemize}

\item \textit{\boldmath $z^{m}_{n}$ is the $n^{th}$ output bit of the multiplier implemented with $P_{m}(x)$.}

\item \textit{\boldmath $z^{m_0,m_1}_{n}$ is the $n^{th}$ output bit of the obfuscated multiplier, obfuscated with functions of $P_{m_0}(x)$ and $P_{m_1}(x)$. The \underline{true} polynomial of this multiplier is implemented with $P_{m_0}(x)$.}

\item \textit{\boldmath $\delta ^{r}_{c}$ is the \underline{obfuscation term} in the obfuscated reduction structure at the $r^{th}$ row and $c^{th}$ column.}

\end{itemize}

To obtain the difference between the reduction structures, vector-wise XOR is applied to the two matrices, $M_{0}$ and $M_1$ generated by \textit{structureGen}. The resulting matrix is denoted as $M'$. The positions of the non-zero element in $M'$, \{($r'_i$,$c'_j$), 1 $\leq$ $i,j$ $\leq$ $m$\}, indicate the differences. The obfuscated reduction structure is first created by copying $M_{0}$. The elements at positions ($r'_i$,$c'_j$) are replaced by $\delta ^{r'_i}_{c'_j}$. The function of $\delta$ is defined by Equation \ref{eqn:obf}. In the actual hardware implementation, $p$ is a known constant to the designers and authorized users, namely \textit{dummy switch}. The dummy switch is implemented by introducing camouflaged standard cell introduced in Section \ref{sec:background_cam_cell} during technology mapping process. 

\vspace{-2mm}
\begin{equation} \label{eqn:obf}
\small
\delta ^{r'_i}_{c'_j} = M_{0}(r'_i,c'_j) \cdot p + M_{1}(r'_i,c'_j) \cdot \bar{p}
\vspace{-1mm}
\end{equation}

\textbf{Example 1}: We illustrate the obfuscation process using the two GF($2^4$) multiplications shown in Figure \ref{fig:obf_4_example}. The resulting multiplier performs multiplication with $P_{0}(x)$ and is obfuscated with the multiplication with $P_{1}(x)$. $M'$ is created by XORing $M_0$ and $M_1$, which includes seven non-zero elements at positions $({r'_i}, {c'_j})$ = \{(2,1), (3,1), (3,2), (2,3), (4,3), (3,4), (4,4)\}. Thus, seven $\delta ^{r'_i}_{c'_j}$ are required for this obfuscation. The obfuscation terms of Figure 4-(c) are shown in Equation \ref{eqn:obf}. The obfuscated multiplication structure is first created by copying $M_0$, and then is updated by replacing the elements at $({r'_i}, {c'_j})$ = \{(2,1), (3,1), (3,2), (2,3), (4,3), (3,4), (4,4)\} with $\delta ^{r'_i}_{c'_j}$ in Equation \ref{eqn:obfuscation_term}. The output function is XORing all the terms at each column in the final step. For example, the MSB in the obfuscated multiplier ($z^{01}_{3}$) is computed as $s_3$ $\oplus$ $\delta ^{2}_{1}$ $\oplus$ $\delta ^{3}_{1}$ $\oplus$ $s_6$. The rest of the logic in the obfuscated multiplier remains the same as in any GF($2^4$) multiplier.

\vspace{-2mm}
\begin{equation} \label{eqn:obfuscation_term}
\small
\vspace{-1mm}
\begin{split}
\delta ^{2}_{1} = s_4\cdot p + 0\cdot\bar{p};~\delta ^{3}_{1} = s_5\cdot p + 0\cdot\bar{p};\\
\delta ^{3}_{2} = 0\cdot p + s_5\cdot\bar{p};~\delta ^{2}_{3} = 0\cdot p + s_4\cdot\bar{p};\\
\delta ^{4}_{3} = s_6\cdot p + 0\cdot\bar{p};~\delta ^{3}_{4} = s_5\cdot p + 0\cdot\bar{p};\\
\delta ^{4}_{4} = s_6\cdot p + 0\cdot\bar{p};~~~~~~~~~~~~~~~~~~~~~~~
\end{split}
\end{equation}

An iterative obfuscation approach is applied to generate obfuscated multiplier with more than two functions. With performing obfuscation with three or more functions, the designer must choose the irreducible polynomial for the true function (e.g. $P_0(x)$), and also choose the \underline{order of obfuscation} among the other functions. For example, consider a scenario in which the designer wants to design an obfuscated GF($2^4$) multiplier with three functions to replace the multiplier block in the ECC hardware, with one more irreducible polynomial $P_{2}(x)$=$x^4$+$x^3$+$x^2$+$x^1$+1 (c.f. Table \ref{tbl:2_to_4_px}). 
The true polynomial is $P_{0}(x)$. Let $M_{0}$, $M_{1}$, and $M_{2}$ be the multiplication structures of $P_{0}(x)$,$P_{1}(x)$, and $P_{2}(x)$. If the order is $P_{1}(x)$$\rightarrow$$P_{2}(x)$, our approach first generates the intermediate obfuscated structure $M'$ with inputs $M_{0}$ and $M_{1}$, and the generates the finalized design by obfuscating $M'$ with $M_{2}$. We can see that 1) in order to obfuscate $n+1$ functions, the number of obfuscation iterations is $n$; 2) the maximum number of functions in one GF($2^m$) multiplier is limited by the total number of irreducible polynomials that have degree $m$. For this iterative approach, the size of the final multipliers are effected by the order of obfuscations. This occurs because the total number and complexity of the obfuscation terms ($\delta$) vary with across the different orders. This has been further explored in Section \ref{sec:permutation_test}.

\vspace{-3mm}
\subsection{Optimization}

Two optimization techniques are introduced to reduce the overhead of the obfuscation approach, 1) early constant propagation and 2) obfuscation term reduction.

\subsubsection{Early constant propagation} It turns out that there could exist a large number of obfuscation terms generated have zero entries. For example, in Equation \ref{eqn:obfuscation_term}, all $\delta ^{r'_i}_{c'_j}$ have constant zero. In which case, those terms can be reduced from AND-OR logic into simple AND functions. 

\subsubsection{Obfuscation term reduction} Two types of reduction are introduced: \textit{a) merging the equivalent $\delta$ terms.} For example, $\delta ^{3}_{1}$ and $\delta ^{3}_{4}$ will be merged since they have the same functionality; \textit{b) reducing non-equivalent $\delta$ terms.} Two $\delta$ in the same column can be merged if one $\delta$ is $s_x$$\cdot$$p$, and the other is $s_y$$\cdot$$\bar{p}$, $x$$\neq$$y$. For example, in Figure 4-(c), in the third column ($z_1$), $\delta ^{2}_{3}$ and $\delta ^{4}_{3}$ can be replaced by $\delta_{reduce}$=$s_6$$\cdot$$p$+$s_4$$\cdot$$\bar{p}$. This removes one term in the third column, which reduces one XOR function for $z_1$ by introducing one OR function. This is because XOR is a more complex Boolean function than OR.


\begin{figure*}[t]
\centering
\vspace{-22mm}
\begin{minipage}{0.42\textwidth}
  \centering
\includegraphics[width=1\textwidth]{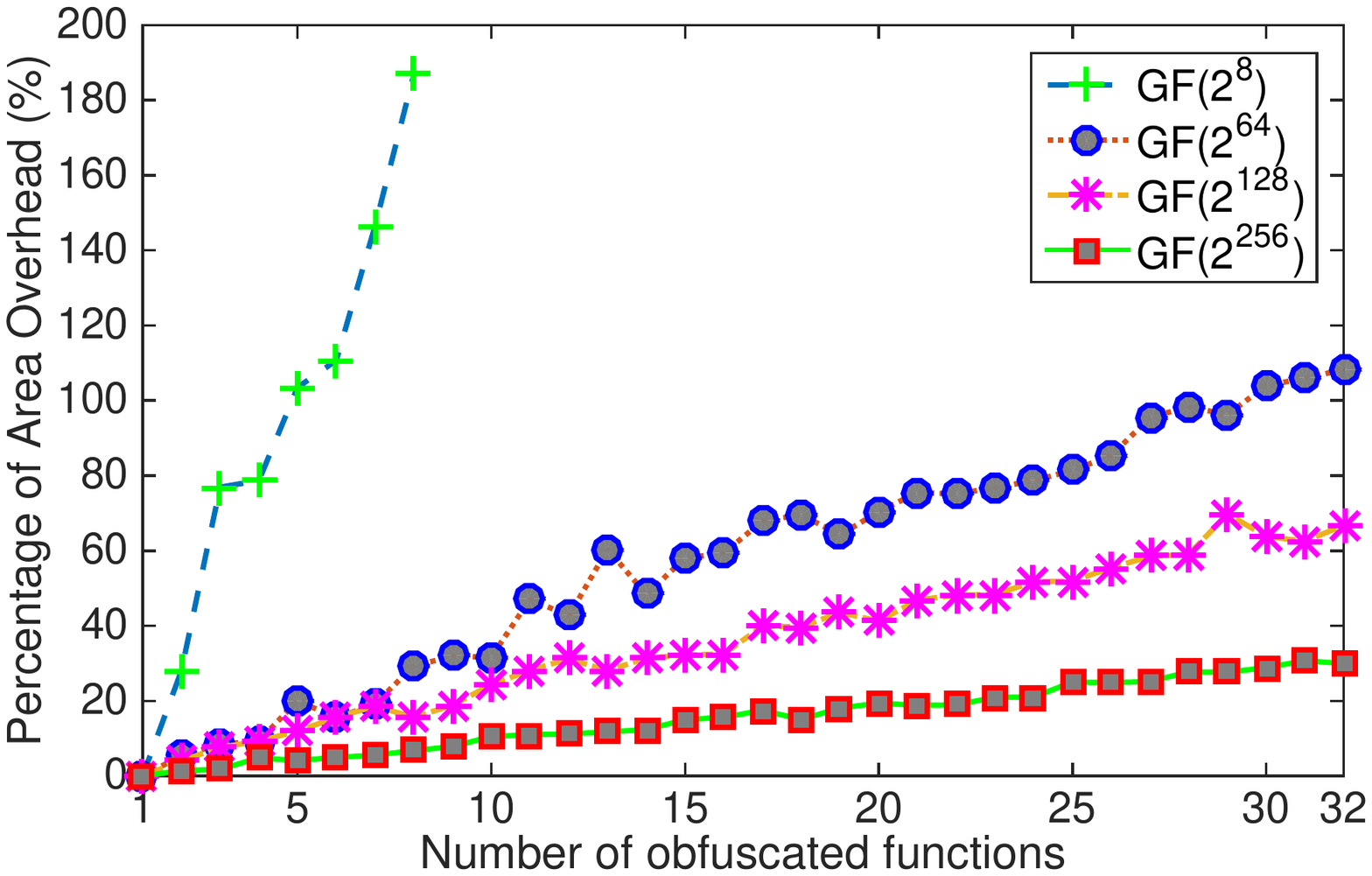}
\vspace{-32mm}
\subcaption{Area overhead vs. number of obfuscated functions.}\label{fig:1a}
\end{minipage}%
\begin{minipage}{0.42\textwidth}
  \centering
\includegraphics[width=1\textwidth]{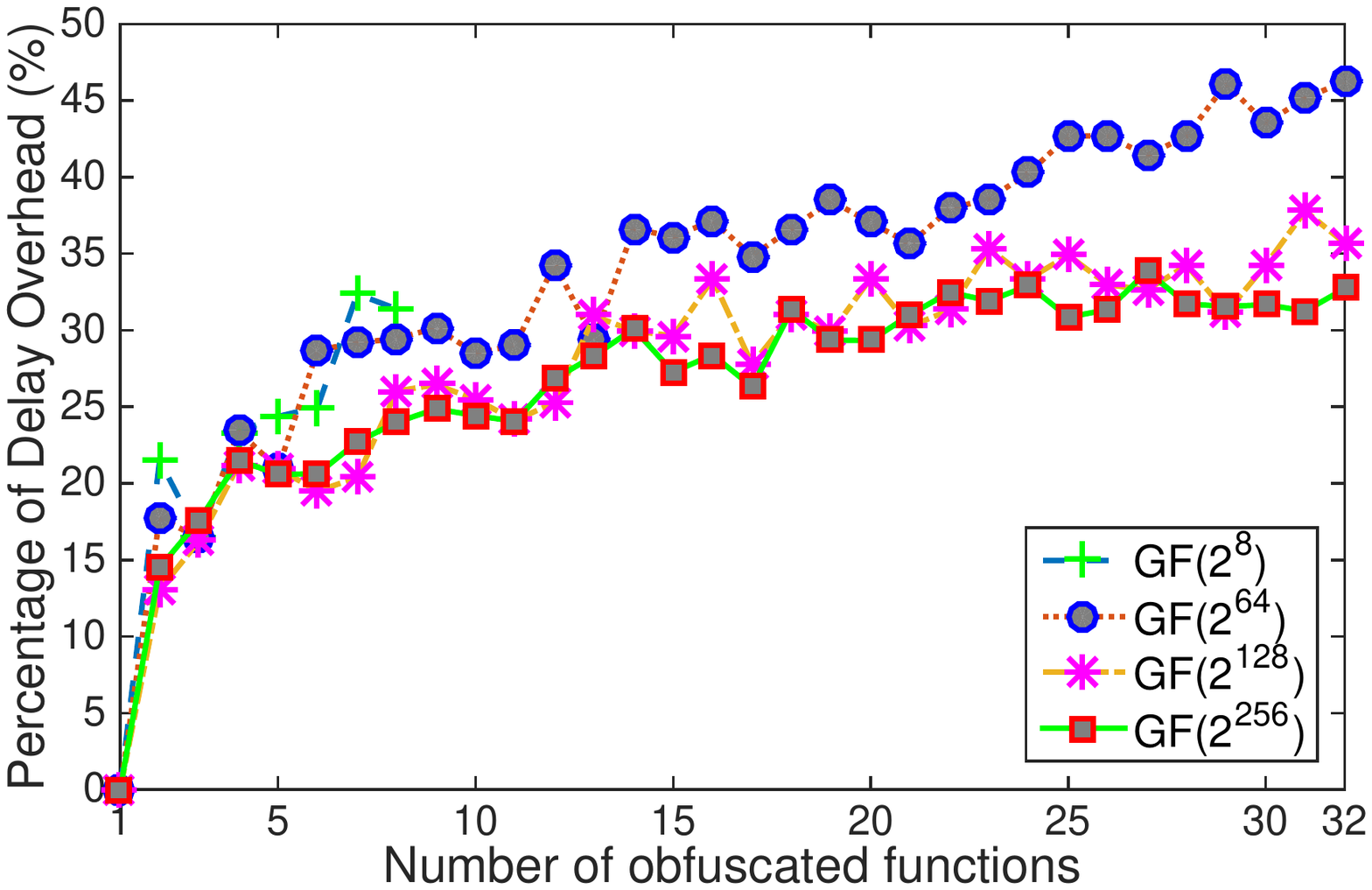}
\vspace{-32mm}
\subcaption{Delay overhead vs. number of obfuscated functions.}\label{fig:1b}
\end{minipage}%
\caption{Area and Delay overhead analysis with $m$ = 8, 64, 128, 256, up to 32 functions obfuscated.}
\vspace{0mm}
\label{fig:result_all}
\end{figure*}

\begin{figure*}[t]
\centering
\vspace{-27mm}
\begin{minipage}{0.4\textwidth}
  \centering
\includegraphics[width=1\textwidth]{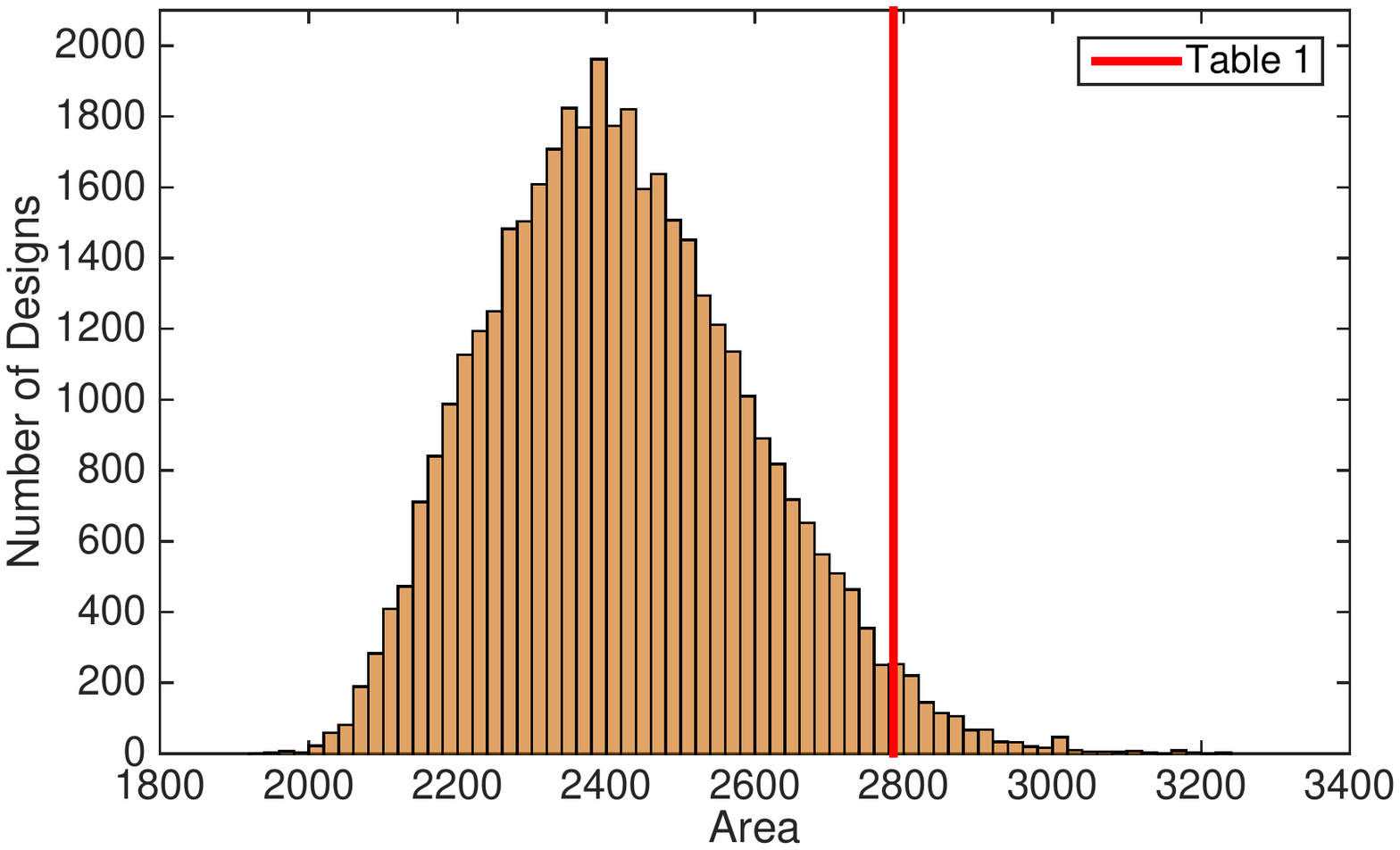}
\vspace{-32mm}
\subcaption{Distribution of area cost.}\label{fig:1a}
\end{minipage}%
\begin{minipage}{0.4\textwidth}
  \centering
\includegraphics[width=1\textwidth]{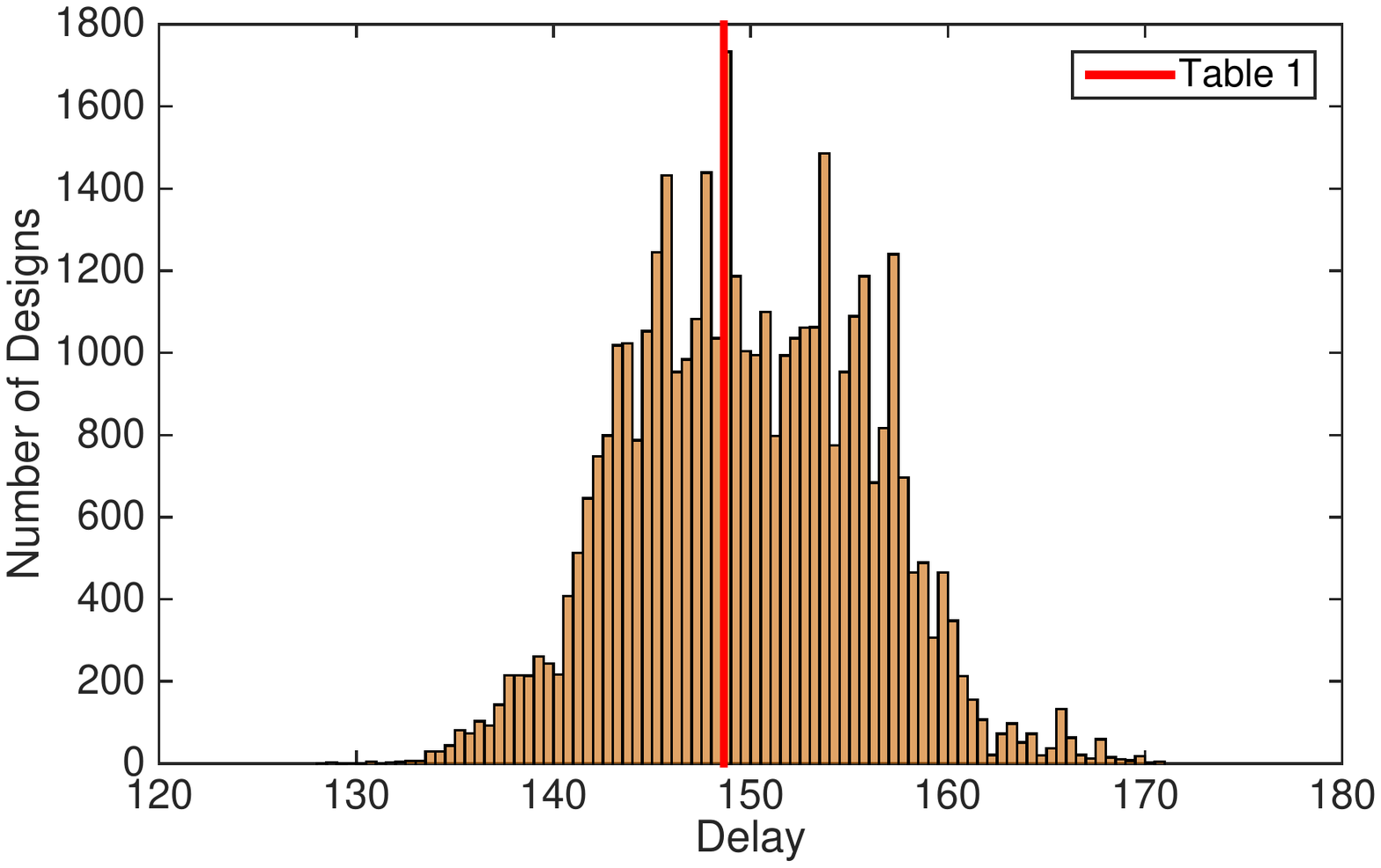}
\vspace{-32mm}
\subcaption{Distribution of delay.}\label{fig:1b}
\end{minipage}%
\caption{Design cost of the GF($2^8$) multipliers with eight function obfuscated with exhaustive permutations of the obfuscation orders, i.e., 40320 designs in total.}
\vspace{-4mm}
\label{fig:permutation}
\end{figure*}



\vspace{-2mm}
\section{Experimental results}\label{sec:result}

The proposed approach is evaluated by creating obfuscated Galois Field multipliers with various numbers of viable $GF(2^m)$ multiplications. The designs generated by our approach are mapped using the open source synthesis tool ABC \cite{abc-link}, with a 14nm technology library. The bit-width $m$ varies from 8 to 256. The irreducible polynomials are obtained from \cite{poly_table}. The runtime of generating obfuscated multipliers is not included in this section because all runtimes are less than one second. In Table \ref{tbl:all_results}, we can see that obfuscating 16 functions for $m$=\{64,128,256\} requires 60\%$\times$, 30\%$\times$, and 20\%$\times$ area overhead. The delay overheads are 36\%, 30\%, and 25\%. On average, the cost of adding an extra obfuscated function is 1.8\% area and delay. The number of obfuscated functions for GF($2^8$) is limited to 8 because there are only eight primitive irreducible polynomials in this field.



\vspace{-3mm}
\subsection{Design Cost Analysis}

To further analyze the design cost, we evaluate the \underline{total} area and delay overhead with the number of obfuscated functions from 2 to 32, with $m$=8, 64, 128, and 256. The x-axis shows the number of obfuscated functions, and the y-axis represents the overhead of area/delay. In Figure \ref{fig:result_all}, we can see that:
\begin{itemize}

\item the area overhead increases almost linearly with the number of functions increasing; on average, the cost of adding an extra obfuscated function is 1.8\% area and delay.

\item {given the same number of obfuscated functions, the area overhead and the delay overhead decrease as $m$ (bit-width of the multiplier) increasing. For example, the area and delay overhead of obfuscating eight functions for GF($2^8$) multiplier are 186\% and 33\%; for GF($2^{256}$), they are 8\% and 22\%.  This shows that our approach advances in obfuscating large Galois Field arithmetic applications.} 

\item the overheads occasionally decrease when the number of obfuscated functions increases. This is because the obfuscation terms $\delta$ introduced may become the \textit{don't care} logic, which helps the technology mapping process to improve the results \cite{DBLP:conf/fpga/MishchenkoBJJ09}.  

\end{itemize}


\vspace{-2mm}
\subsection{Order of Obfuscations}\label{sec:permutation_test}

As mentioned in Section \ref{sec:obfuscation}, the size of the obfuscated multipliers are affected by the orders of the iterative obfuscations. The main reason is that using different orders, the number of $\delta$ and the complexity of these $\delta$ can be very different. An exhaustive permutation study over GF($2^8$) is shown in Figure \ref{fig:permutation} to demonstrate the impact of the obfuscation order. All possible eight-function obfuscated GF($2^8$) multipliers are generated by the proposed approach, while each order corresponds to one permutation of \{$P_{0}$, $P_{1}$, ..., $P_{7}$\}. Thus, the total number of designs in Figure \ref{fig:permutation} is 8!=40320. The results are collected by ABC with 14nm technology library. The x-axis shows the area/delay, and the y-axis shows the number of designs in a given range of area/delay. The area varies from 1800-3300, and the delay ranges from 125-170. We can see that the order of obfuscations has great impact on the design cost of the obfuscated multipliers. Comparing the result to the order used in Table \ref{tbl:all_results} (\textit{Table 1} in Figure \ref{fig:permutation}), area=2786 and delay=148.64, that design can be further improved by exploring the choice of orders. The future work will focus on finding the good orders for efficient obfuscation using machine learning.

\subsection{Evaluation of Attacks}\label{sec:attack}

We apply the SAT-based attack technique using the two tools released publicly \cite{subramanyan-15}\cite{yu2017incremental}. The inputs to the tools are Verilog design with extra syntax for defining the de-camouflaging problems. We develop a set of camouflaged GF circuits using the proposed approach, including 8-bit, 12-bit,16-bit and 32-bit GF functions. Each of these circuits includes four camouflaged GF functions. Regarding the BDD approach \cite{xu2017novel}, we measure the performance of constructing the BDDs of the camouflaged circuit using the same CUDD package \cite{somenzi:2009-cudd}. The results are shown in Table \ref{tbl:attack}. The SAT-based attack techniques cannot obtain the true function with only three dummy functions after 16-bit within 12 hours. BDD construction fails at 16-bit as well due to the memory explosion. Note that the cryptograph applications such as ECC could have large GF operators. 

\begin{table}[!htb]
\centering
\caption{Evaluations of the attack techniques \cite{subramanyan-15}\cite{yu2017incremental}\cite{somenzi:2009-cudd} in reverse engineering the true functionality of the camouflaged GF multipliers. Time out limit is 12 hours.}
\begin{tabular}{|l|r|r|r|r|}
\hline
\multicolumn{1}{|c|}{bit-width} & \multicolumn{1}{c|}{\cite{subramanyan-15}} & \multicolumn{1}{c|}{\cite{yu2017incremental}} & \multicolumn{1}{c|}{BDD} & \multicolumn{1}{c|}{\cite{yu:2016-tcad-verification}}\\ \hline
8                               & 4 s                            & 6 s                            & 1 s            & 4 s           \\ \hline
12                              & 33 s                           & 29 s                           & 4 s             & MO         \\ \hline
16                              & \textgreater12 hrs             & \textgreater12 hrs             & \textgreater 16 GB   & MO    \\ \hline
\end{tabular}
\label{tbl:attack}
\end{table}

\vspace{-2mm}
\section{Conclusion}
\vspace{-2mm}

In this paper, we introduce an obfuscation approach over Galois Field, mainly focusing on obfuscation GF multiplications. Our approach generates GF multipliers with multiple irreducible polynomials obfuscated, to prevent the actual irreducible polynomial being reverse engineered. A complete design methodology is developed and evaluated with a set of GF multipliers, with up to 32 functions obfuscated. The results show that our approach can obfuscate the GF multipliers with low overhead in design performance. We also evaluate the strength of obfuscation over Galois Field using SAT-based and BDD-based techniques. The future work will focus on leveraging machine learning algorithms in searching the best obfuscation order(s). 

\vspace{-2mm}

\bibliographystyle{IEEEtran}
\bibliography{security,verification_ycunxi,synthesis,dan}
\end{document}